\documentclass[%
 reprint,
superscriptaddress,
amsmath,amssymb,
aps,
pra,
12pt,
onecolumn,
]{revtex4-2}
\usepackage{moresize}
\usepackage{setspace}
\usepackage{graphicx}
\usepackage{dcolumn}
\usepackage{bm}
\usepackage[colorlinks=true,linkcolor=blue]{hyperref}%
\usepackage{tikz,siunitx}
\usetikzlibrary{shapes.geometric,shapes.symbols}
\hypersetup{
    colorlinks=true,
    linkcolor=blue,
    citecolor=blue,
    filecolor=magenta,      
    urlcolor=blue,
}

\begin{document}

\preprint{APS/123-QED}

\title{Metamorphosis of transition to periodic oscillations in a turbulent reactive flow system}

\author{Beeraiah Thonti}
\author{Sivakumar Sudarsanan}
\author{Ramesh S. Bhavi}
\author{Anaswara Bhaskaran}
\author{Manikandan Raghunathan}
\author{R. I. Sujith}
\affiliation{%
  Department of Aerospace Engineering, Indian Institute of Technology Madras, Chennai 600 036, India}%
\affiliation{
  Centre of Excellence for Studying Critical Transition in Complex Systems, Indian Institute of Technology Madras, Chennai 600 036, India}

\date{\today}
\makeatletter
\gdef\@ptsize{0} 
\makeatother
\doublespace
\begin{abstract}
The emergence of periodic oscillations is observed in various complex systems in nature and engineering. Thermoacoustic oscillations in systems comprising turbulent reactive flow exemplify such complexity in the engineering context, where the emergence of oscillatory dynamics is often undesirable. In this work, we experimentally study the transition to periodic oscillations within a turbulent flow reactive system, with varying fuel-to-air ratio, represented by equivalence ratio as a bifurcation parameter. Further, we explore the change in the nature of the transition by varying a secondary parameter. In our system, we vary the thermal power input and the location of the flame stabilizer position individually as a secondary parameter. Our findings reveal five qualitatively distinct types of transitions to periodic oscillations. Two types of these transitions exhibit a continuous nature. Another two types of transitions involve multiple shifts in the dynamical states consisting of both continuous and discontinuous bifurcations. The last type of transition is characterized by an abrupt bifurcation to high-amplitude periodic oscillations. Understanding this metamorphosis of the transition - from continuous to discontinuous nature - is critical for advancing our comprehension of the dynamic behavior in turbulent reactive flow systems. The insights gained from this study have the potential to inform the design and control of similar engineering systems where managing oscillatory behavior is crucial.
\end{abstract}

\maketitle


\section{Introduction} \label{introduction}
Natural and engineering systems are often complex systems, consisting of several interacting subsystems. Complex systems undergo transitions from one state to another state with a variation in a system parameter. Often, these transitions lead to self-sustained periodic oscillations. Such self-organized oscillations emerge when positive feedback is established among the subsystems of a complex system \citep{scheffer2012anticipating,angeli2004detection}. The emergence of these oscillations in complex systems has been a subject of long-standing research, particularly because they can often be detrimental to the functioning of the system.\\
\hspace*{10mm}Turbulent reactive flows are complex systems comprising subsystems such as acoustic, hydrodynamic and heat release rate fields \citep{chu1958non, tandon2023multilayer}. In these systems, large amplitude self-organized oscillations emerge due to positive feedback between the acoustic and the heat release rate oscillations \citep{lieuwen2005combustion}. Such oscillatory behavior is referred to as thermoacoustic instability, which hinders the development of gas turbine and rocket engines \citep{juniper2018sensitivity, lieuwen2005combustion}. These high-amplitude oscillations necessitate emergency shutdowns in land-based power-generating gas turbine engines, while in aircraft engines decrease the size of the safe operation regime. The vibrations generated due to the occurrence of high-amplitude oscillations cause fatigue and wear and tear and, in severe cases, lead to structural failure \citep{lieuwen2005combustion}. In rocket engines, self-organized oscillations intensify the heat transfer overwhelming the thermal protection system. These oscillations can also affect the onboard electronics, and damage the guidance and navigation systems, leading to failure or delays in space programs \citep{sujith2021thermoacoustic}. Understanding the dynamics of transitions to self-sustained oscillations is crucial for engineers to design safer and more efficient combustion systems. Furthermore, the emergence of these oscillations in turbulent reactive flows has piqued the interest of the nonlinear dynamics community. These systems exhibit intriguing dynamical behaviors, including chimera states \citep{mondal2017onset}, multifractality \citep{nair2014multifractality, raghunathan2020multifractal}, and the formation of spatial patterns \citep{george2018pattern, sujith2020complex}.\\
\hspace*{10mm}Traditionally, the transition to combustion instability is modeled as a supercritical or subcritical Andronov-Hopf bifurcation \citep{clavin1994turbulence, lieuwen2005combustion} (often referred to simply as Hopf bifurcation). A bifurcation is a qualitative change in the system behavior due to a slight change in a system parameter beyond a critical value, and the parameter is referred to as a bifurcation parameter \citep{Strogatz2019-pb}. In the Hopf bifurcation, when the bifurcation parameter varies beyond a critical value, a stable fixed point loses its stability as the sign of the real part in a pair of complex conjugate eigenvalues changes, leading to limit cycle oscillations. The bifurcation parameter value at which the fixed point loses its stability is called the Hopf point. Similarly, in thermoacoustic systems, a transition to oscillatory behavior emerges from a stable operation when the bifurcation parameter exceeds a critical value.\\
\hspace*{10mm}Hopf bifurcation accurately describes the transition from stable operation to thermoacoustic instability in laminar combustors, which manifests as a bifurcation from a fixed point to a limit cycle. However, it is crucial to acknowledge that the paradigm of Hopf bifurcation may not be applicable to transitions in all thermoacoustic systems. In particular, turbulent combustors may exhibit a more intricate transition to limit cycle oscillations, involving dynamical states that extend beyond the scope of the classical Hopf bifurcation framework.\\
\hspace*{10mm}\citet{lieuwen2002experimental} described the transition from a stable operation to an unstable operation as a noisy supercritical or subcritical bifurcation from a fixed point to a limit cycle in turbulent reactive flow systems.  In a turbulent combustor, conducting determinism tests on acoustic pressure fluctuations during the stable operation (also referred to as combustion noise in the parlance of thermoacoustics) revealed that the fluctuations are deterministic and are characterized as high dimensional chaos contaminated with noise \citep{nair2013loss, tony2015detecting}. Nair \textit{et al.} \citep{nair2014intermittency} showed that turbulent reactive flows undergo a transition from combustion noise to thermoacoustic instability via a state of intermittency with an increase in Reynolds number $(Re)$ as the control parameter. The state of intermittency is characterized by epochs of high-amplitude periodic oscillations amidst epochs of low-amplitude aperiodic fluctuations. This transition occurred in a continuous manner, with the dominant peak in the amplitude spectrum (FFT) increasing continuously.\\
\hspace*{10mm}Recent studies have also reported abrupt transitions to thermoacoustic instability in turbulent combustors. In a turbulent combustor with preheated inlet air, \citet{pawar2021effect} observed that a transition from stable operation to thermoacoustic instability occurs with a corresponding significant jump in the root mean square (RMS) value of acoustic pressure oscillations. They decreased the fuel-to-air ratio, quantified using the term equivalence ratio as the bifurcation parameter. In a turbulent annular combustor, \citet{singh2021intermittency} observed a transition from chaos to high-amplitude periodic oscillations, with varying equivalence ratio, at high bulk flow velocity. The transition consists of primary supercritical bifurcation to a low amplitude thermoacoustic instability, followed by a secondary abrupt transition to high amplitude thermoacoustic instability. Extending this finding, \citet{bhavi2023abrupt} demonstrated similar dynamics in other combustors and employed a reduced order model with stochastic noise to explain the observations. Recently, Pavithran \textit{et al.} \citep{pavithran2023tipping} conducted experiments in a turbulent combustor, varying $Re$ in a continuous manner. They found that the system remained at low amplitude aperiodic acoustic pressure oscillations for slow variations of $Re$, while fast variations caused an abrupt transition to high-amplitude periodic oscillations. Building upon this study, \citet{joseph2024explosive} further characterized this abrupt transition as an explosive synchronization transition.\\
\hspace*{10mm}Research spanning various fields shows that various dynamical systems undergo both continuous and discontinuous transitions. In ecology, the transitions from a minimal population state to a state of population outbreak is observed as a continuous or discontinuous transition for the populations of spruce budworms \citep{jones1975application}, forest caterpillars \citep{rose1981ecological}, and butterflies \citep{loehle1989catastrophe}. In quantum mechanics, the nature of switching of laser polarization, whether continuous or discontinuous, depends on the choice of the appropriate control parameter and its scanning range \citep{zhang2008polarization}. The nature of dynamo transition in magneto-hydrodynamic turbulence changes from supercritical to subcritical with a decrease in the magnetic Prandtl number \citep{verma2013supercriticality}. In dynamical systems, the nonlinearities in the system impacted by the variation in an additional parameter cause a change in the nature of the transition, which is referred to as a change in criticality \citep{marsden2012hopf}.\\
\hspace*{10mm}To understand this change in criticality in thermoacoustic systems, \citet{etikyala2017change} conducted a systematic experimental study on a horizontal Rijke tube. A Rijke tube is a simple open-open cylindrical tube consisting of a heat source located at a position of one-quarter of the axial length of the tube. The authors increased the heater power as a bifurcation parameter and obtained a transition from a fixed point to limit cycle oscillations. Further, they reported a change in the nature of the transition from supercritical Hopf to subcritical Hopf with increasing airflow rate as a secondary parameter,  attributing this change to the stabilizing and destabilizing influence of the governing nonlinearities in the system. \\
\hspace*{10mm}Building on these insights in laminar systems, we turn our attention to thermoacoustic systems comprising turbulent reactive flows, which are inherently more complex. In these systems, the stable operation manifests as high dimensional chaos and thermoacoustic instability is characterized by limit cycle oscillations. The transition from stable operation to thermoacoustic instability occurs via several dynamical states, including the state of intermittency \citep{nair2014intermittency} and mixed-mode oscillations \citep{singh2021intermittency}. Consequently, the transition to high amplitude periodic oscillations in turbulent combustors manifests as a sequence of bifurcations involving many dynamical states and is significantly more intricate than in the case of laminar systems \citep{sujith2021thermoacoustic}.  Thus, understanding the impact of nonlinearities on the transition to thermoacoustic instability in turbulent reactive flow systems is crucial.\\
\hspace*{10mm}Towards this purpose, we conduct experiments in a turbulent reactive flow system with a bluff body employed to stabilize the flame, referred to as flame stabilizer. We vary two sets of pairs of parameters, namely $(1)$ fuel-to-air ratio, represented by equivalence ratio ($\phi$) and thermal power input ($\mathcal{P}$) and $(2)$ $\phi$ and the location of flame stabilizer position ($x_f$). We decrease the value of $\phi$ at a fixed  $\mathcal{P}$ to observe the transition from aperiodic fluctuations to high amplitude periodic oscillations. We perform these experiments at various values of $\mathcal{P}$, keeping the location of $x_f$ constant, acquiring a set of transitions. Similarly, we decrease the value of $\phi$ at a particular location of $x_f$ to observe a transition from aperiodic fluctuations to high amplitude periodic oscillations. We perform these experiments at various locations of $x_f$, keeping the value of $\mathcal{P}$ as constant, acquiring another set of transitions. In summary, we varied the equivalence ratio as a bifurcation parameter causing the transition, with the value of $\mathcal{P}$ and location of $x_f$ as secondary parameters influencing the governing nonlinearities, thereby impacting the transition dynamics. We analyze the time series of acoustic pressure oscillations ($p'$) to characterize the dynamical states at various values of bifurcation parameters. We discover an intriguing \textbf{metamorphosis of the dynamical transition} from chaos to order in turbulent reactive flow systems on varying a secondary parameter along with the bifurcation parameter.\\
\hspace*{10mm}The rest of the paper is structured as follows: Section~\ref{experiments} provides an overview of the experimental setup and the measurement systems used in this study. In Section~\ref{results}, we discuss the detailed analysis of the time series of acoustic pressure fluctuations during a set of transitions to periodic oscillations at various values of $\mathcal{P}$ and also another set of transitions to periodic oscillations at various values of $x_f$. We discuss the scope for future work and summarize the conclusion in Section~\ref{discussion}.\\
\section{Experimental setup}\label{experiments}
\begin{figure}[ht!]
\begin{center}
\includegraphics[scale=0.75]{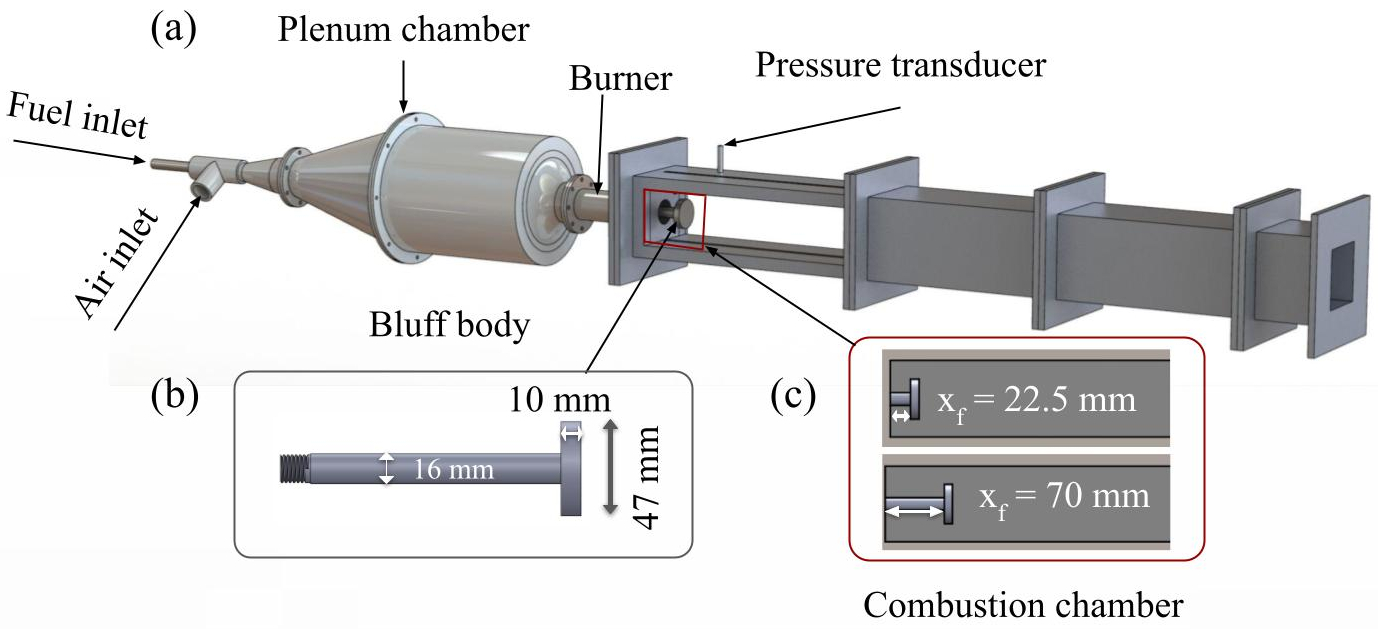}
\caption{(a) Schematic of the experimental configuration of the turbulent combustor, (b) illustrates the dimensions of the bluff body, and (c) depicts the extreme positions of the bluff body ($x_f$) employed in this experimental study in relation to the backward-facing step of the combustor.} 
\label{BB_tara}
\end{center}
\end{figure}

We perform experiments in a turbulent combustor with a bluff body as a flame stabilizer, as depicted in figure~\ref{BB_tara} (a). The experiments were conducted under atmospheric conditions. The setup comprises three primary components: the plenum chamber, followed by the burner, leading into the combustion chamber. The plenum chamber minimizes the flow fluctuations in the air entering through the inlet. The combustion chamber has a square cross-section of $90$ mm $\times$ $90$ mm in size and a length of $1100$ mm. A quartz glass of $400$ mm $\times$ $90$ mm $\times$ $10$ mm is fixed to one side of a combustor section to provide optical access. The bluff body (BB) is a circular disc of $47$ mm diameter and $10$ mm thick and is mounted on a hollow shaft of $16$ mm outer diameter (Fig.~\ref{BB_tara}(b)). The flame stabilizer position from the backward-facing step can be adjusted, as depicted in figure~\ref{BB_tara}(c).\\ 
\hspace*{10mm}Liquified petroleum gas (LPG) consisting of $60\%$ butane $+$ $40\%$ propane by volume is used as the fuel. LPG is injected radially through four holes of $1.7$ mm diameter into the air in the burner, resulting in a technically premixed air-fuel mixture entering the combustion chamber. The mass flow rates of fuel and air are measured in standard liter per minute (SLPM) and are controlled with mass flow controllers (Alicat Scientific, MCR $2000$ SLPM series for air, MCR $100$ SLPM for fuel). The mass flow controllers have a measurement uncertainty of $\pm$($0.8$ $\%$ of reading $+$ $0.2$ $\%$ of full scale). In the experiments, the mass flow rate of the fuel is varied from $1.03\pm 0.02$ g/s to $1.69 \pm 0.02$ g/s, and the mass flow rate of air is varied from $8.54\pm 0.27$ g/s  to $21.92 \pm 0.25$ g/s. Therefore, the uncertainty in the equivalence ratio is $\pm 0.02$. Acoustic pressure oscillations in the combustion chamber are measured using a piezoelectric pressure transducer (PCB103B02) with a sensitivity of $0.2306$ mV/Pa and a measurement uncertainty of $\pm 0.15$ Pa. The pressure transducer is flush mounted on the combustor wall at a location $40$ mm from the backward-facing step of the combustor.\\
In our experiments, we fix the mass flow rate of fuel and increased the mass flow rate of air, consequently decreasing the global equivalence ratio ($\phi$). $\phi$ is calculated based on the fuel-air mixture entering the combustion chamber. We vary the value of $\phi$ in the range of $2.16-1.26$ in a quasi-static manner. We perform the experiments at various thermal power input $(\mathcal{P})$ values ranging from $57-93$ kW and various values of flame stabilizer position ranging from $70-22.5$ mm from the backward-facing step of the combustor. At each value of $\phi$, we operate the combustor for a duration of $5$ s, then acquire the time series of acoustic pressure oscillations for $3$ s at $20000$ samples per second. We vary the equivalence ratio in steps of $0.09$ during the early stage of the transition ($\phi \geq 1.69$), where the system is away from the onset of periodic oscillations. As the system approaches the onset of oscillations ($\phi \leq 1.64$), we refine the variation of $\phi$ to a step size of $0.05$. We perform experiments for each transition a minimum of three times to ensure repeatability.
\section{Results }\label{results}
We observe a transition from aperiodic fluctuations to periodic oscillations by decreasing the global equivalence ratio ($\phi$) as a bifurcation parameter in a quasi-static manner. We perform these experiments by varying a secondary parameter. At each specific value of the secondary parameter, we have a transition from chaos to periodic oscillations obtained with a decrease in $\phi$. \\
\newcommand{\tikzcircle}[2][red,fill=red]
    {\tikz[baseline=-0.5ex]\draw[#1,radius=#2] (0,0) circle ;}
    \newcommand{\coloredbox}{\textcolor{violet}{\rule{0.5em}{0.5em}}}
    \newcommand{\tikzsymbol}[2][circle]{\tikz[baseline=-0.5ex]\node[shape=#1,draw,#2]{};}%
    \newcommand{\fillednabla}{\tikz{\node[fill=blue, minimum size=1ex] {$\nabla$};}}
    \newcommand{\hollowbox}[1]{\textcolor{#1}{\fboxsep=0pt\fbox{\phantom{\rule{0.5em}{0.5em}}}}}
\begin{figure*}
     \includegraphics[scale =0.45]{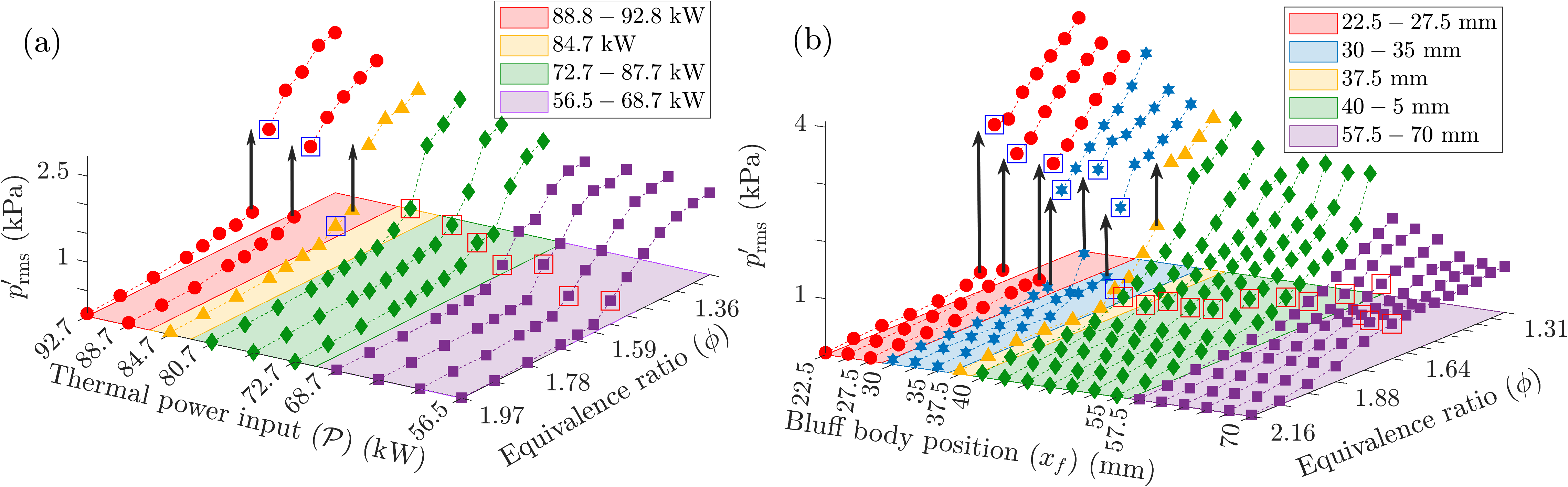}
    \caption{Variation of root mean square (RMS) values of acoustic pressure ($p'$) oscillations as a function of the global equivalence ratio ($\phi$) during the set of transitions to periodic oscillations: (a) with increasing value of thermal power input ($\mathcal{P}$) (value of $x_f$ fixed at $30$ mm) and (b) with decreasing the value of flame stabilizer position ($x_f$) (value of $\mathcal{P}$ fixed at $80.7$ kW). The subset of transitions is categorized based on similarity in the variation of $p'_\mathrm{rms}$ and color-coded accordingly. Various markers such as $\coloredbox, \tikzsymbol[diamond]{scale=0.5, fill=green}$, \textcolor{yellow}{$\blacktriangle$}, $\color{blue}\star$ and $\tikzcircle[red, fill=red]{2.5pt}$ are used to distinguish among the subset of transitions. The onset of oscillations is highlighted during the continuous transitions with a red box ($\hollowbox{red}$) and the discontinuous transition with a blue box ($\hollowbox{blue}$). We observe that the nature of the transition changes from continuous to discontinuous as we increase the value of $\mathcal{P}$ or decrease the value of $x_f$ as a secondary parameter.}
    \label{3d_map}
\end{figure*}
\hspace*{10mm}Varying the secondary parameter causes a switch in the nature of the transition from continuous to discontinuous \citep{kuehn2021universal, etikyala2017change}. In our experiments, we observe both the continuous and discontinuous transitions to periodic oscillations, depending on the value of the secondary parameter of the system. In our study, we calculate the RMS value of the acoustic pressure oscillations ($p^\prime$) as a measure to characterize the nature of the transition. During a continuous transition, the measures calculated from the time series of observables, such as RMS values or dominant peak from the amplitude spectrum (FFT), vary continuously. In contrast, these measures vary abruptly during a discontinuous transition.\\
\hspace*{10mm}Figure~\ref{3d_map} is rich in information, however, not detailed. We will walk through it with the reader. In Section~\ref{Mf_route}, we include a detailed analysis of $p'$ during the distinct transitions at various values of $\mathcal{P}$. We reconstructed the high-dimensional phase space by analyzing the pressure fluctuation data ($p'$) for a duration of $0.5$ s using the time delay embedding theorem \citep{10.1007/BFb0091924}. For this purpose, we calculated the delay time ($\tau$) using the average mutual information (AMI) method \citep{fraser1986independent} and determined the embedding dimension with the false nearest neighbors (FNN) technique \citep{kennel1992determining}. 
\subsection{Change in the nature of the transition to periodic oscillations with thermal power input} \label{Mf_route}
In this set of experiments, we vary the global equivalence ratio $(\phi)$ as the bifurcation parameter and thermal power input $(\mathcal{P})$ as a secondary parameter. Here, we fix the location of $x_f$ at $30$ mm. We perform experiments at various values of $\mathcal{P}$ ranging from $56.6$ kW to $92.8$ kW in approximate steps of $4$ kW. Consequently, we obtain a set of transitions to periodic oscillations at different values of $\mathcal{P}$, as depicted in figure~\ref{3d_map}\textcolor{blue}{a}. Transitions within this range are segregated into four distinct subsets based on the qualitative similarity in the variation of $p'_\mathrm{rms}$. Different color codes and markers are used to distinguish the subsets. We specifically discuss each transition from the four subsets (Fig.~\ref{states_28}, ~\ref{states_36}, ~\ref{states_42} and ~\ref{states_46}). We plot the variation of $p'_{\mathrm{rms}}$ as a function of $\phi$ to represent the transition from a regime of aperiodic fluctuations to periodic oscillations.
\subsubsection{Type $1$ transition: C-I-SNA-NLC route of continuous transition to periodic oscillations}\label{56.5kw}
\begin{figure}
     \centering
    \includegraphics[width= 8.5 cm,height= 8.5 cm]{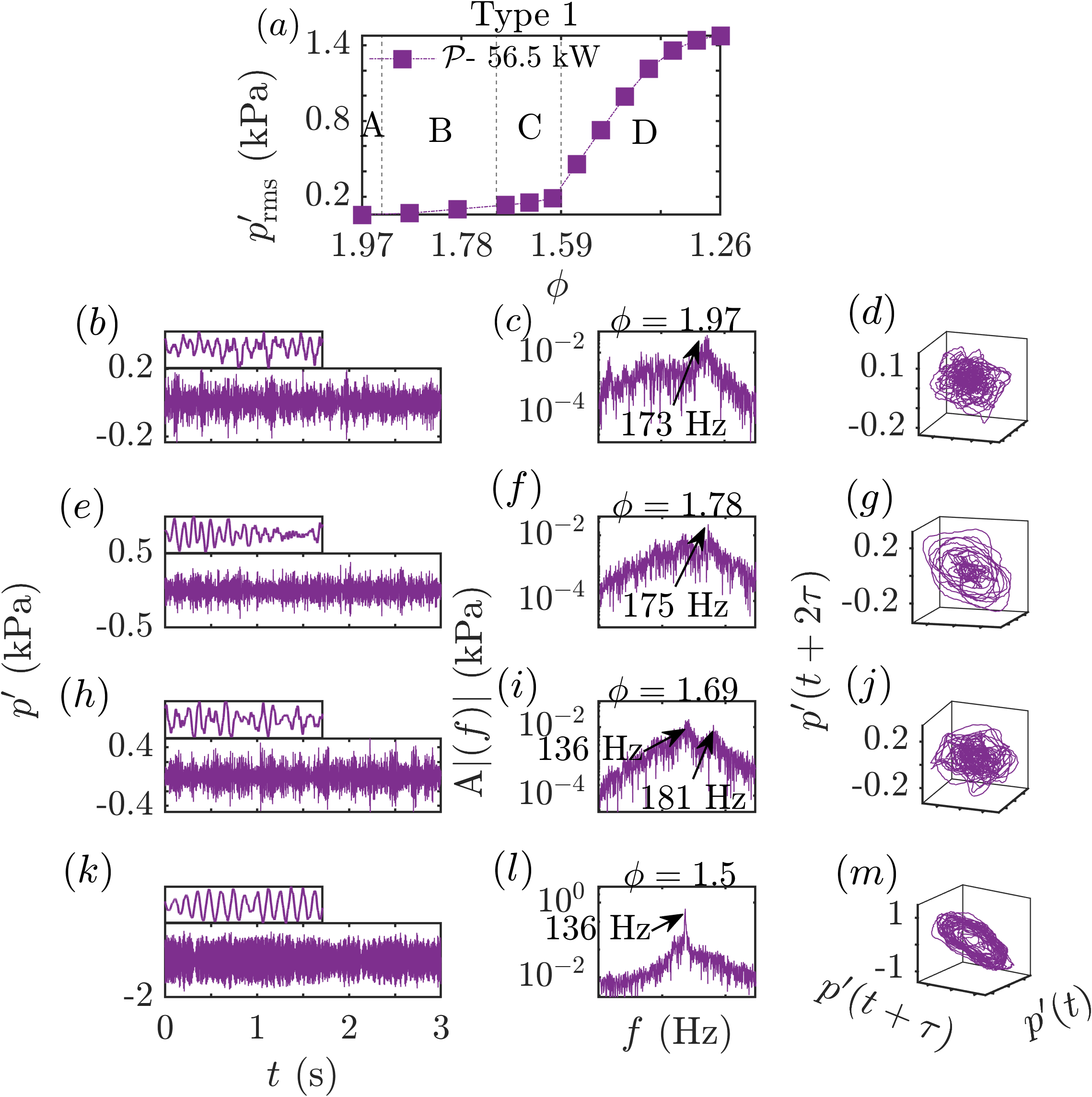}
    \caption{\textbf{Type $1$: C-I-SNA-NLC route of continuous transition}. (a) Variation of $ p'_\mathrm{rms}$ as a function of $\phi$ during the transition to noisy limit cycle oscillations. Various dynamical regimes are marked with A-D. The time series, amplitude spectrum, and the reconstructed phase space of $p'$ during the states of (b-d) chaos, (e-g) intermittency, (h-j) SNA, and (k-m) noisy limit cycle.}
    \label{states_28}
 \end{figure}

\hspace*{10mm} We observe a continuous transition to periodic oscillations when the values of $\mathcal{P}$ are fixed between $56.5$ kW and $68.7$ kW. These transitions are represented by ($\coloredbox$) and shaded in violet in figure~\ref{3d_map}\textcolor{blue}{a}. For the value of $\mathcal{P}$ of $56.5$ kW, during the transition, we observe a gradual increase in $p^\prime_{\mathrm{rms}}$ (Fig.~\ref{states_28}\textcolor{blue}{a}). The dynamical states observed during this transition are the states of chaos (C), intermittency (I), strange nonchaotic attractor (SNA), and noisy limit cycle (NLC). These states are indicated with A, B, C, and D in figure~\ref{states_28}\textcolor{blue}{a}, respectively.\\

\begin{figure}
    \centering
    \includegraphics[width= 8.5 cm,height= 10 cm]{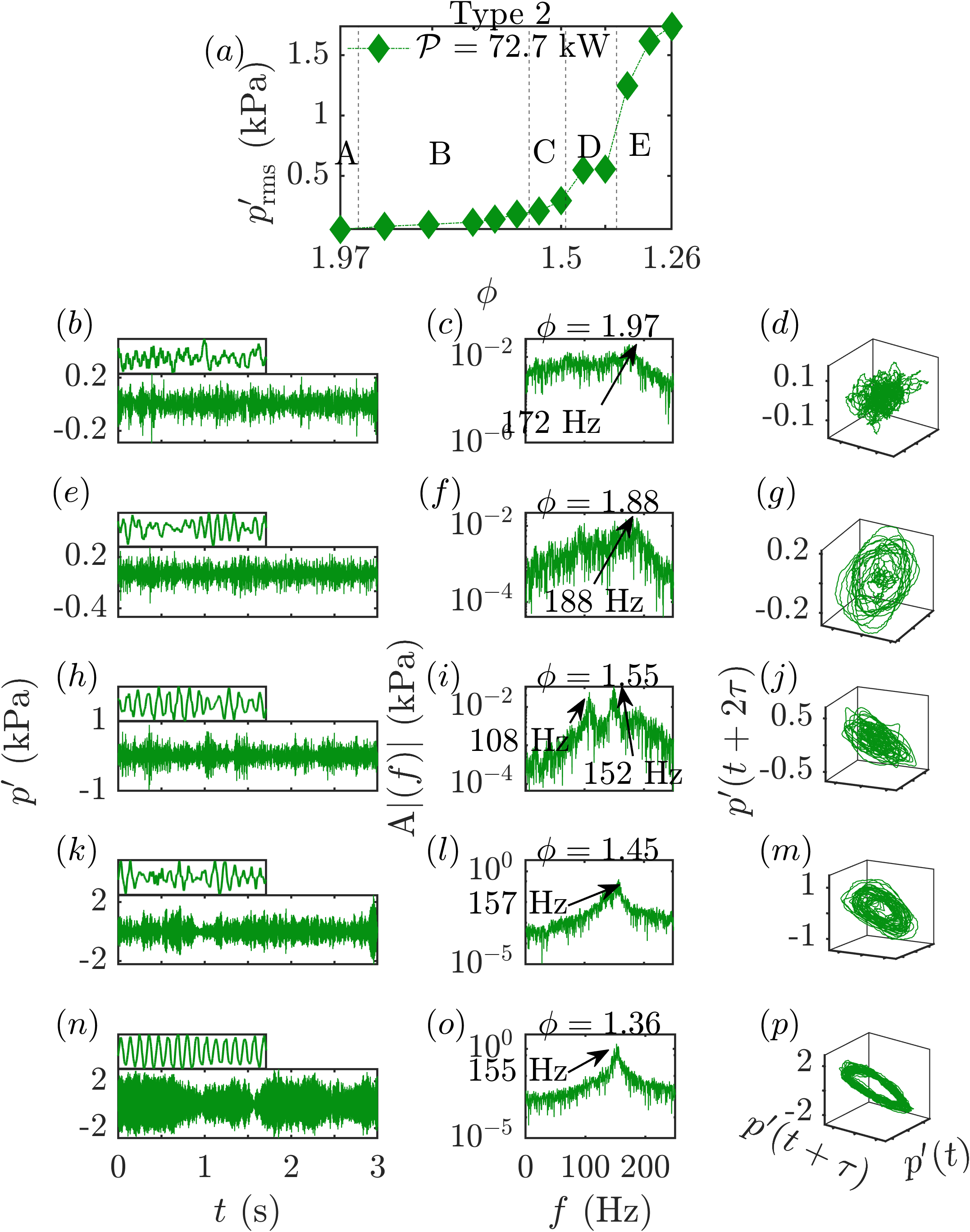}
    \caption{\textbf{Type $2$: C-I-SNA-NLC-CLC route of continuous transition}.(a) Variation of $ p'_\mathrm{rms}$ as a function of $\phi$ during the transition to clean limit cycle. Various dynamical regimes are marked with A-E. The time series, amplitude spectrum, and the reconstructed phase space of $p'$ during the states of (b-d) chaos, (e-g) intermittency, (h-j) SNA, (k-m) noisy limit cycle, and (n-p) clean limit cycle.}
    \label{states_36}
\end{figure}

\begin{figure}
    \centering
    \includegraphics[width= 8.5 cm,height= 10 cm]{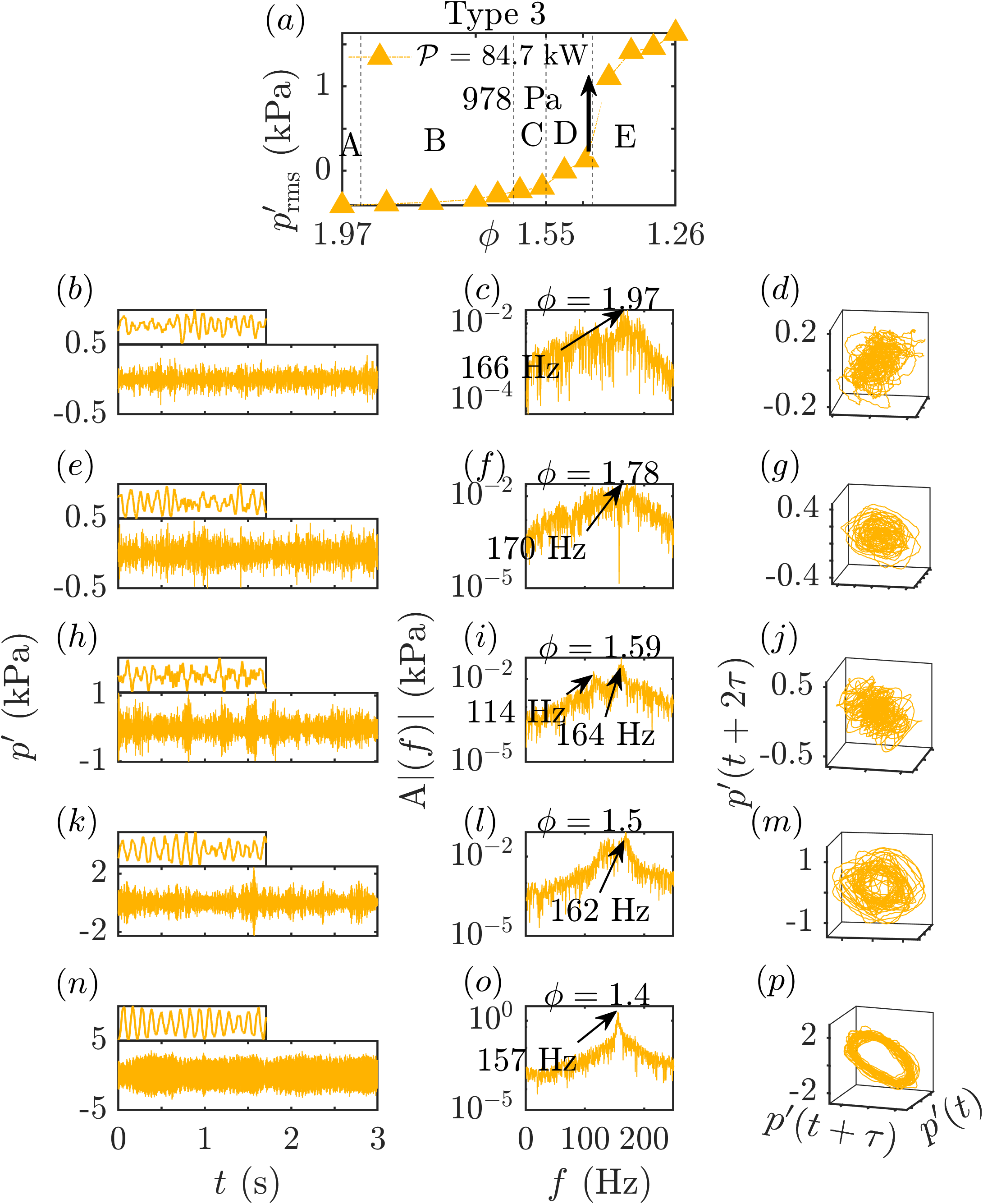}
    \caption{\textbf{Type $3$: C-I-SNA-NLC-CLC route of primary continuous bifurcation to NLC followed by a secondary discontinuous bifurcation to CLC}. (a) Variation of $p'_\mathrm{rms}$ as a function of $\phi$ during the transition to clean limit cycle. Various dynamical regimes are marked with A-E. The time series, amplitude spectrum, and the reconstructed phase space of $p'$ during the states of (b-d) chaos, (e-g) intermittency, (h-j) SNA, (k-m) noisy limit cycle and (n-p) clean limit cycle.}
    \label{states_42}
\end{figure}

\hspace*{10mm}We observe the state of chaos characterized by low amplitude aperiodic fluctuations \citep{tony2015detecting, nair2013loss} at a value of $\phi = 1.97$ (Fig.~\ref{states_28}\textcolor{blue}{b}). The amplitude spectrum of $p'$ exhibits a broad peak around $173$ Hz, which closely aligns with the theoretical quarter-wave mode frequency of the combustor duct ($f = c/4L$, where $c$ is the speed of sound and $L$ is the length of the combustor duct), estimated to be $159$ Hz (Fig.~\ref{states_28}\textcolor{blue}{c}). The average temperature of the flame is measured to be $1200 \pm 9$ K, resulting in a speed of sound of $700 \pm 4.5$ m/s ($c = \sqrt{\gamma RT}$ where $\gamma$ is the adiabatic index, and $R$ is the gas constant). The trajectory in the reconstructed phase space is noisy and cluttered (Fig.~\ref{states_28}\textcolor{blue}{d}).\\ 
\hspace*{10mm}As we decrease the value of $\phi$ to $1.78$, we observe the state of intermittency characterized by epochs of high amplitude periodic oscillations amidst epochs of low amplitude aperiodic fluctuations (Fig.~\ref{states_28}\textcolor{blue}{e}). The amplitude spectrum has a broad peak at $175$ Hz (Fig.~\ref{states_28}\textcolor{blue}{f}). The phase space reconstruction for this state reveals the trajectory switching between high amplitude periodic and low amplitude aperiodic behavior, indicating that the system is alternately switching between these two states (Fig.~\ref{states_28}\textcolor{blue}{g}).\\
\hspace*{10mm}At the value of $\phi$ of $1.69$, we observe that the behavior of $p'$ (Fig.~\ref{states_28}\textcolor{blue}{h}) and the trajectory in the reconstructed phase space (Fig.~\ref{states_28}\textcolor{blue}{j}) appears qualitatively similar to chaotic state. However, the amplitude spectrum reveals two broad peaks around $136$ Hz and $181$ Hz (Fig.~\ref{states_28}\textcolor{blue}{i}), which is different from the chaotic state. Motivated by these observations, the author of this paper conducted multiple advanced tests, including $0-1$ test, correlation dimension test, and singular continuous spectrum analysis and confirmed that this state is characterized as a state of strange nonchaotic attractor (SNA) \cite{thonti2024strangenonchaoticattractorunforced}. The state of SNA has a fractal structure, similar to chaos. However, do not exhibit sensitivity to initial conditions making it nonchaotic in nature.\\
\hspace*{10mm}A further decrease in the value of $\phi$ below $1.59$ reveals the emergence of periodic oscillations with amplitude modulations characterized as noisy limit cycle. The noisy limit cycle generally possesses the fundamental features of a clean limit cycle, such as the deterministic nature and regularity. However, the presence of noise disrupts the smoothness and the predictability of oscillations. At the value of $\phi$ of $1.5$, the time series appears periodic with amplitude modulations (Fig.~\ref{states_28}\textcolor{blue}{k}), and the corresponding amplitude spectrum shows a peak at $136$ Hz (Fig.~\ref{states_28}\textcolor{blue}{l}). Furthermore, the trajectory of the phase space forms a thick ring (Fig.~\ref{states_28}\textcolor{blue}{m}). The thickness of the ring is larger compared to a clean limit cycle (Fig.~\ref{states_36}\textcolor{blue}{p}, ~\ref{states_42}\textcolor{blue}{p} and ~\ref{states_46}\textcolor{blue}{m}) due to the deviation in the trajectory from the mean trajectory of the limit cycle during a few acoustic pressure cycles. The observed deviations in the trajectory may be attributed to phase jitter \citep{doi:10.1260/175682709789141528} and seen in form of variation of pressure oscillations. This noise, which potentially arises from the intricate interaction between the intensity of turbulence and the position of the bluff body, plays a significant role in the dynamics of the system during the transition.  

\subsubsection{Type $2$ transition: C-I-SNA-NLC-CLC route of continuous transition to periodic oscillations}\label{72.7kw}
\hspace*{10mm}When we increase the value of $\mathcal{P}$ to the range of $72.7$ and $80.7$ kW, the continuous nature of the transition persists as shown in figure~\ref{3d_map}\textcolor{blue}{a}, where the respective subset of transitions are plotted with ($\tikzsymbol[diamond]{scale=0.5, fill=green}$) and shaded in green. Notably, $p^\prime_{\mathrm{rms}}$ attains higher values at $\phi = 1.26$ compared to $p^\prime_{\mathrm{rms}}$ corresponding to same value of $\phi$ during the previous subset of transitions. During the transition to periodic oscillations at the value of $\mathcal{P}$ of $72.7$ kW, the system traverses through various dynamical states in a sequence: the states of chaos (C), intermittency (I), SNA, noisy limit cycle (NLC) and finally to clean limit cycle (CLC), indicated the regions with A-E in figure~\ref{states_36}\textcolor{blue}{a} respectively.\\
\hspace*{10mm}The dynamics observed in the time series (Fig.~\ref{states_36}\textcolor{blue}{b, e, h}, and \textcolor{blue}{k}), amplitude spectrum (Fig.~\ref{states_36}\textcolor{blue}{c, f, i}, and \textcolor{blue}{l}) and reconstructed phase space (Fig.~\ref{states_36}\textcolor{blue}{d, g, j}, and \textcolor{blue}{m}) during the dynamical states of chaos, intermittency, SNA and noisy limit cycle are qualitatively similar to the dynamical states observed during type $1$ transition as described in Section~\ref{56.5kw}. The values of dominant frequencies during the respective states are mentioned in figure~\ref{states_36}(\textcolor{blue}{c, f, i, l}). The additional dynamical state attained in this subset is a state of clean limit cycle, during which a dominant frequency of $155$ Hz (Fig.~\ref{states_36}\textcolor{blue}{o}). The trajectory in the reconstructed phase space forms a thin ring structure (Fig.~\ref{states_36}\textcolor{blue}{p}). During the transition discussed in the previous section, the route to periodic oscillations is C-I-SNA-NLC (Fig.~\ref{states_28}\textcolor{blue}{a}). While, for this transition, the route to periodic oscillations is C-I-SNA-NLC-CLC (Fig.~\ref{states_36}\textcolor{blue}{a}).
\subsubsection{Type $3$ transition: C-I-SNA-NLC-CLC route of primary continuous bifurcation followed by a discontinuous bifurcation to periodic oscillations}\label{84.7kw}
During the previous subset of transitions discussed, the entire transition occurs in a continuous manner. Subsequently, when we increase the values of $\mathcal{P}$ to $84.7$ kW, the nature of the transition changes from a continuous one to one that consists of a primary continuous bifurcation to noisy limit cycle, followed by a secondary discontinuous bifurcation to a clean limit cycle (the specific transition is plotted with (\textcolor{yellow}{$\blacktriangle$}) and highlighted in yellow in figure~\ref{3d_map}\textcolor{blue}{a}). The results corresponding to $\mathcal{P} = 84.7$ kW are shown in figure~\ref{states_42}. Interestingly, in spite of a change in the nature of the transition, the route to periodic oscillations remains similar to type $2$ transition. However, during type $2$ transition, when the system shifts from NLC to CLC, the dominant frequency remains nearly constant. In contrast, during type $3$ transition, the frequency is shifting. Initially, a decrease in the value of $\phi$ causes a gradual increase in  $p^\prime_{\mathrm{rms}}$. Correspondingly, the system shifts from the state of chaos to the state of intermittency and then to the state of SNA. With a further decrease in the value of $\phi$, the system shifts to a noisy limit cycle, and $p^\prime_{\mathrm{rms}}$ increases gradually. Decreasing the value of $\phi$ below a critical value causes the system to shift to a clean limit cycle with an abrupt jump in $p^\prime_{\mathrm{rms}}$ of $978$ Pa. The dynamics during each dynamical state are qualitatively similar, as explained in Section~\ref{72.7kw}. The values of the dominant frequency during each state are mentioned in the figure~\ref{states_42}(\textcolor{blue}{c, f, i, l, o}). This transition, which consists of a primary continuous bifurcation followed by a secondary discontinuous bifurcation, aligns with the findings of \citet{bhavi2023abrupt}.
\begin{figure}
    \centering
    \includegraphics[width= 8.5 cm,height= 8.5 cm]{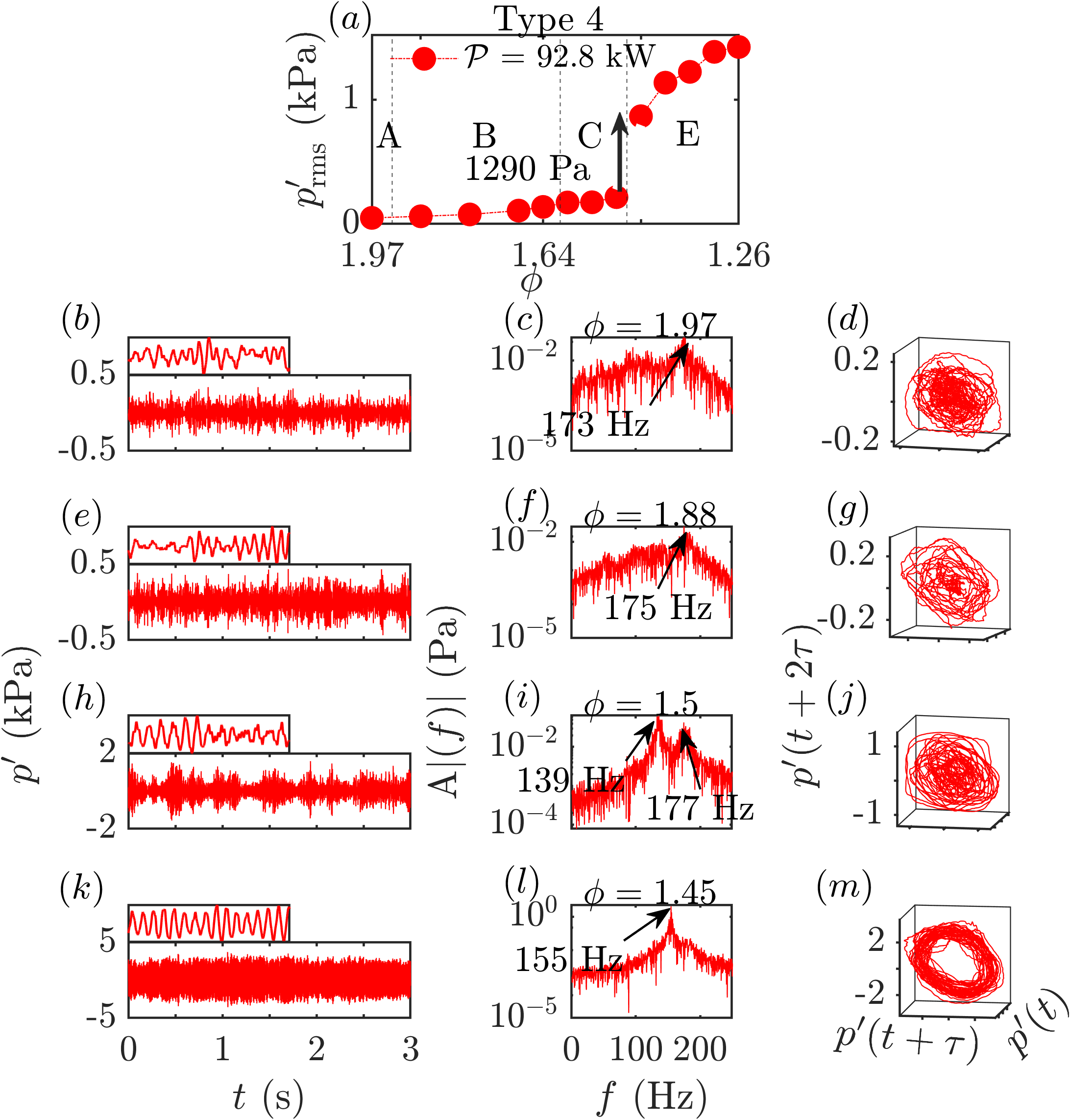}
    \caption{\textbf{Type $4$: C-I-SNA-CLC route of discontinuous transition}.(a) Variation of $ p'_\mathrm{rms}$ as a function of $\phi$ during the transition to clean limit cycle. Various dynamical regimes are color-shaded and marked with A-E. Time series, amplitude spectrum, and the reconstructed phase space of $p'$ during the states of (b-d) chaos, (e-g) intermittency, (h-j) SNA, and (k-m) clean limit cycle oscillations.}
    \label{states_46}
\end{figure}
\subsubsection{Type $4$ transition: C-I-SNA-CLC route of discontinuous transition to periodic oscillations}\label{92.8kW}
\hspace*{10mm}We observe the discontinuous transition to periodic oscillations With a further increase in the value of $\mathcal{P}$ to $88.8$ kW and $92.8$ kW. This subset of transitions is shown in figure~\ref{3d_map}\textcolor{blue}{a}, where the specific subset of transitions is represented with ($\tikzcircle[red, fill=red]{2.5pt}$) and highlighted in red. Within this subset of transitions, a decrease in the value of $\phi$ leads to an increase in the value of $p^\prime_{\mathrm{rms}}$ in a gradual manner, followed by an abrupt jump in $p^\prime_{\mathrm{rms}}$ at a critical value of $\phi$. At a value of $\mathcal{P}$ of $92.8$ kW, decreasing the value of $\phi$ initially results in a very gradual increase in $p^\prime_{\mathrm{rms}}$ and the system shifts from the state of chaos to the state of SNA via the state of intermittency (Fig.~\ref{states_46}\textcolor{blue}{a}). With a further decrease in the value of $\phi$, an abrupt jump of $1290$ Pa in the value of $p^\prime_{\mathrm{rms}}$ is observed, with the system shifting to a clean limit cycle. Interestingly, the system bypasses the state of NLC and directly shifts to CLC during this transition.\\
\hspace*{10mm}A summary of the results discussed in the above section is presented in Table \ref{table_mf}. In the following Section~\ref{route to TAI BB}, we discuss the set of transitions obtained at various locations of $x_f$.\\
\begin{table*} [ht]
  \caption{\label{table_mf}Summary of the set of transitions obtained for the values of $\mathcal{P}$ ranges between $56.6$ kW and $92.8$ kW}
\begin{ruledtabular}
\begin{tabular}{lccccr}
\textbf{Type ($\mathcal{P}$ (kW))} & \textbf{No. of  shifts} & \textbf{route of the transition} & \textbf{$p'_{\mathrm{rms}}$}\\
\hline
Type $1$ ($56.6 $) &  $3$ & C$\xrightarrow[\text{}]{\text{cont.}}$ I$\xrightarrow[\text{}]{\text{cont.}}$ SNA $\xrightarrow[\text{}]{\text{cont.}}$ NLC      & \includegraphics[width=1.5 cm,height=0.7 cm]{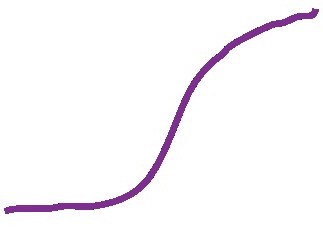} \\ 
Type $2$ ($72.7$) &  $4$ & C $\xrightarrow[\text{}]{\text{cont.}}$ I$\xrightarrow[\text{}]{\text{cont.}}$ SNA$\xrightarrow[\text{}]{\text{cont.}}$ NLC $\xrightarrow[\text{}]{\text{cont.}}$ CLC  & \includegraphics[width=1.5 cm,height=0.7 cm]{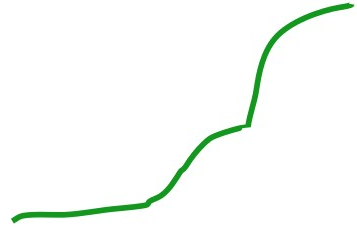}\\ 
Type $3$ ($84.7$)        & $4$ &C$\xrightarrow[\text{}]{\text{cont.}}$ I$\xrightarrow[\text{}]{\text{cont.}}$ SNA$\xrightarrow[\text{}]{\text{cont.}}$ NLC $\xrightarrow[\text{}]{\text{discont.}}$ CLC  &  \includegraphics[width=1.5 cm,height=0.7 cm]{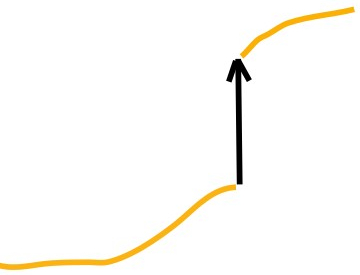}\\ 
Type $4$ ($92.8$)  & $3$ &C$\xrightarrow[\text{}]{\text{cont.}}$ I$\xrightarrow[\text{}]{\text{cont.}}$ SNA $\xrightarrow[\text{}]{\text{discont.}}$ CLC        &\includegraphics[width=1.5 cm,height=0.7 cm]{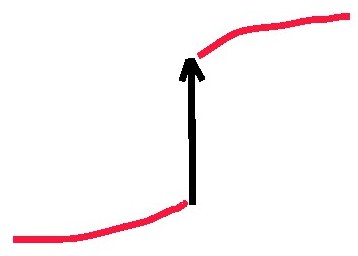}\\ 
\end{tabular}
\end{ruledtabular} 
 {C-chaos}, {I-intermittency}, {SNA-strange nonchaotic attractor}, {NLC-noisy limit cycle}, {CLC-clean limit cycle}, {cont.-continuous transition} and {discont.-discontinuous transition}
\end{table*}

\subsection{Change in the nature of the transition to periodic oscillations with flame stabilizer position} \label{route to TAI BB}
The variation of $p'_{\mathrm{rms}}$ with a decrease in the value of $\phi$ from $2.16$ to $1.31$ at the location of $x_f$ ranging from $70$ to $22.5$ mm relative to the backward-facing step is illustrated in figure~\ref{3d_map}\textcolor{blue}{b}.\\
\begin{figure}
   \centering
    \includegraphics[width=8.5 cm,height=7.5 cm]{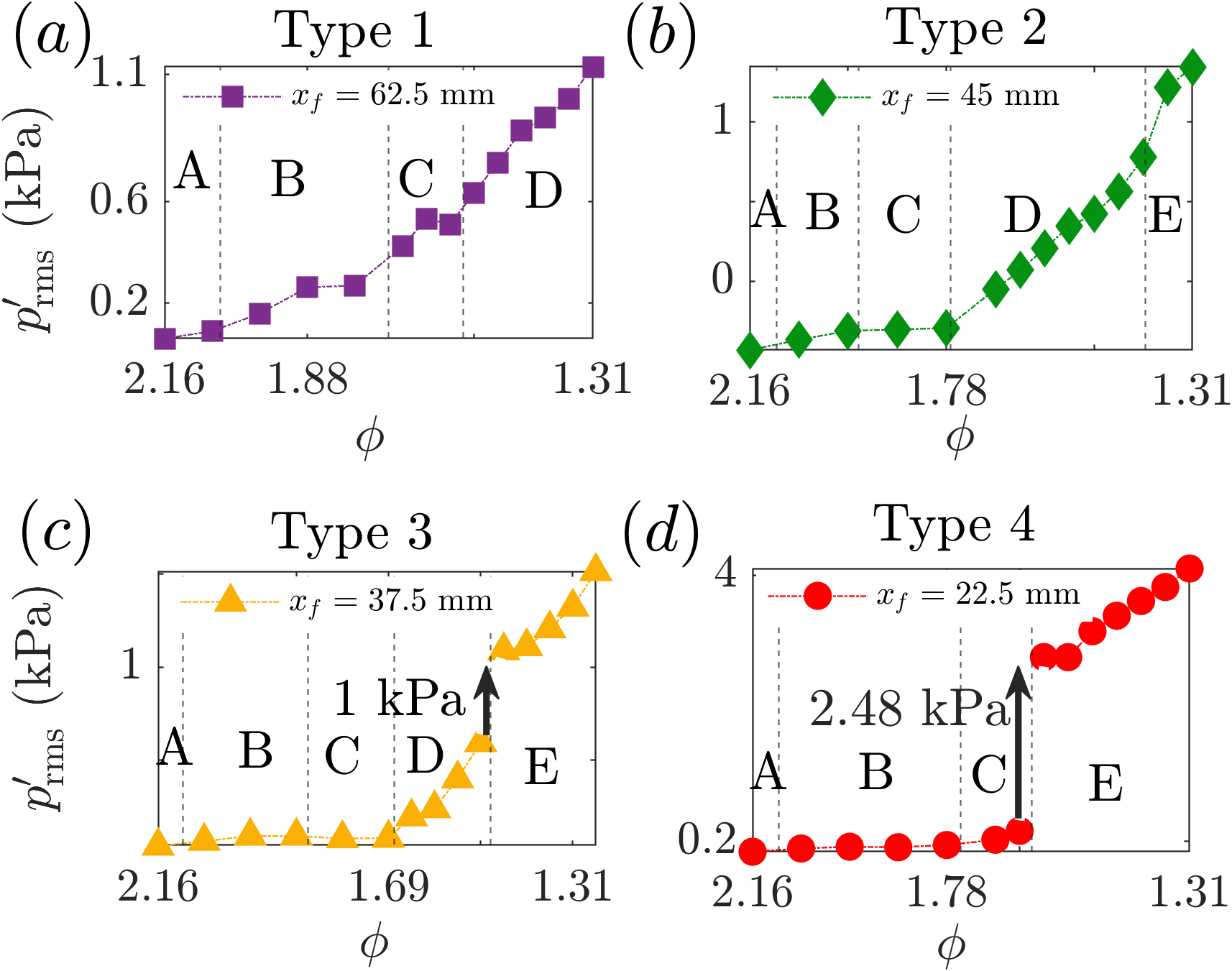} 
    \caption{Variation of $p^\prime_{\mathrm{rms}}$ as a function of $\phi$ during the transition to periodic oscillations for the location of $x_f$ of (a) $62.5$ mm, (b) $45$ mm, (c) $37.5$ mm, and (e) $22.5$ mm, respectively. Arrows highlight abrupt jumps occurring when (c) $x_f=35$ mm, and (d) $x_f=22.5$ mm. The observed dynamical regimes are marked as: A-state of chaos, B-state of intermittency, C-SNA, D-noisy limit cycle, and E-clean limit cycle.}
    \label{bif_bb} 
\end{figure}
\hspace*{10mm}Similar to the set of transitions discussed at various values of $\mathcal{P}$ in Section~\ref{Mf_route}, we also obtain another set of transitions by varying the location of $x_f$ (Fig.~\ref{3d_map}\textcolor{blue}{b}). During this set of transitions, the value of $\mathcal{P}$ is held constant at $80.7$ kW. Similar to the set of transitions discussed previously, we categorize this set into subsets. Beyond the four distinct subsets identified when varying the values of $\mathcal{P}$ as a secondary parameter (Fig.~\ref{3d_map}\textcolor{blue}{a}), we observe an additional type of subset with decreasing value of $x_f$ as a secondary parameter. This new subset of transitions is observed when the value of $x_f$ is between $35$ mm and $30$ mm, which we discuss later. Below, we briefly discuss the transitions from the four subsets that align with those previously identified, and the corresponding bifurcation diagrams are shown in figure~\ref{bif_bb}. A detailed analysis of the time series of $p'$ during the respective transitions is provided in Appendix~\ref{xf_BB}.\\
\hspace*{10mm}When the flame stabilizer is located between $70$ and $57.5$ mm, the transition occurs from a state of chaos to a state of noisy limit cycle oscillations with a gradual increase in $p^\prime_{\mathrm{rms}}$ (Fig.~\ref{3d_map}\textcolor{blue}{b}, with the corresponding subset of transitions are plotted with ($\coloredbox$) and highlighted in violet). Figure~\ref{bif_bb}\textcolor{blue}{a} shows the transition diagram for the value of $x_f$ at $62.5$ mm from this subset of transitions. The route of the transition to periodic oscillations is characterized as C-I-SNA-NLC (region A-D in figure~\ref{bif_bb}\textcolor{blue}{a}) and the nature of the transition is continuous. This transition is qualitatively similar to type $1$ transition (Fig.~\ref{states_28}\textcolor{blue}{a}) discussed in Section~\ref{56.5kw}. 
When the location of $x_f$ is fixed between $55$ mm and $42.5$ mm, the continuous nature of the transition persists. The subset of transitions are represented in figure~\ref{3d_map}\textcolor{blue}{b}, are plotted with ($\tikzsymbol[diamond]{scale=0.5, fill=green}$) and highlighted in green. During this subset of transitions, the occurrence of dynamical states advanced in the bifurcation parameter space, resulting in the system attaining a state of the clean limit cycle. Figure~\ref{bif_bb}\textcolor{blue}{b} shows the transition diagram at $x_f=45$ mm. The route to thermoacoustic instability is characterized as C-I-SNA-NLC-CLC (region A-E in figure~\ref{bif_bb}\textcolor{blue}{b}). This transition is qualitatively similar to type $2$ transition (Fig.~\ref{states_36}\textcolor{blue}{a}) discussed in Section~\ref{72.7kw}.\\
\hspace*{10mm}The continuous nature of the transition is lost when we fix the location of $x_f$ below $40$ mm. When the location of $x_f$ is kept between $40$ mm and $35$ mm, the transition consists of a primary smooth bifurcation to a noisy limit cycle followed by an abrupt bifurcation to a clean limit cycle. In figure~\ref{3d_map}\textcolor{blue}{b}, the subset of transitions is plotted with (\textcolor{yellow}{$\blacktriangle$}) and highlighted in yellow. At a location of $x_f$ of $37.5$ mm, the route of the transition to periodic oscillations is characterized as C-I-SNA-NLC-CLC (region A-E in Fig.~\ref{bif_bb}\textcolor{blue}{c}). Initially, the system shifts from a state of chaos to a state of noisy limit cycle via states of intermittency and SNA with a gradual increase in $p^\prime_{\mathrm{rms}}$ (Fig.~\ref{bif_bb}\textcolor{blue}{c}). Further, decreasing the value of $\phi$ below a critical value, we observe a shift to a clean limit cycle, associated with an abrupt jump in the value of $p^\prime_{\mathrm{rms}}$ (marked with an arrow in figure~\ref{bif_bb}\textcolor{blue}{c}). This transition appears similar to type $3$ transition (Fig.~\ref{states_42}\textcolor{blue}{a}) discussed in Section~\ref{84.7kw}.\\
\hspace*{10mm}Subsequently, when the location of $x_f$ is between $27.5$ and $22.5$ mm, we observe that the nature of the transition changed to discontinuous. In figure.~\ref{3d_map}\textcolor{blue}{b}, the subset of transitions is plotted with ($\tikzsymbol[circle]{scale=0.5, fill=red}$) and highlighted in red. At the value of $x_f$ of $22.5$ mm, the route to periodic oscillations is observed as C-I-SNA-CLC. The dynamical shifts that occur from the states of chaos to SNA via intermittency are associated with a gradual increase in $p^\prime_{\mathrm{rms}}$. With a further decrease in the value of $\phi$ below a critical value, the system directly shifts to a state of a clean limit cycle, skipping the state of a noisy limit cycle. This shift is associated with an abrupt jump of $2480$ Pa in $p^\prime_{\mathrm{rms}}$ as shown in figure~\ref{bif_bb}\textcolor{blue}{d}. The route to periodic oscillations is similar to type $4$ transition (Fig.~\ref{states_46}\textcolor{blue}{a}) discussed in Section~\ref{92.8kW}. We now proceed to analyze a transition in detail that is different from the four types of transitions discussed.
\subsubsection{Type $5$ transition: I-SNA-NLC-CLC route of discontinuous transition to periodic oscillations} \label{35mm}
\begin{figure}
    \centering
    \includegraphics[width=8.5 cm,height= 10 cm]{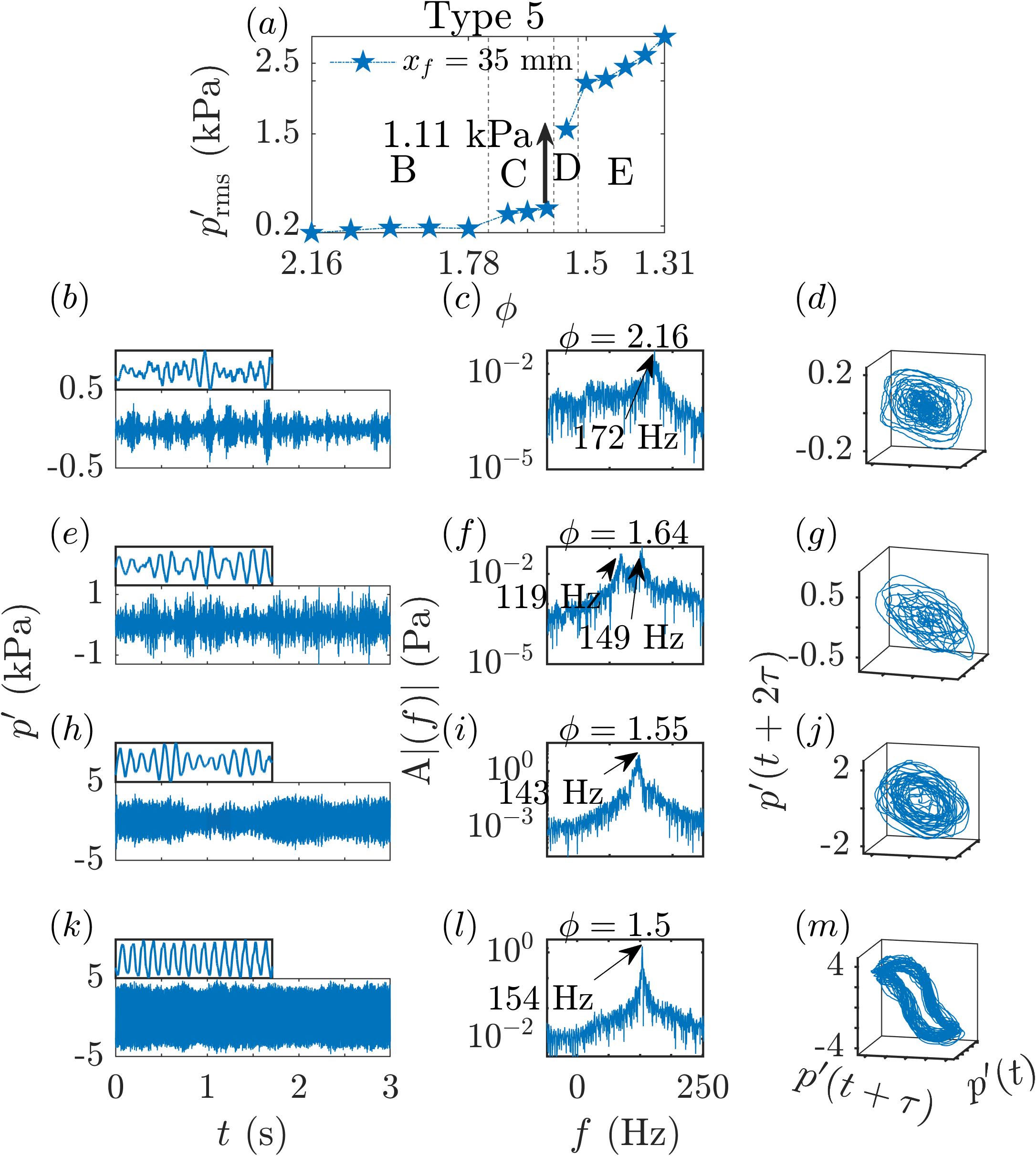} 
    \caption{\textbf{Type $5$: I-SNA-NLC-CLC route of primary abrupt bifurcation to NLC followed by a secondary smooth bifurcation to CLC} (a) Variation of $ p'_\mathrm{rms}$ as a function of $\phi$ during the transition to clean limit cycle. Various dynamical regimes are marked with B-E. Time series, amplitude spectrum, and the reconstructed phase space of $p'$  plots during the states of (b-d) intermittency, (e-g) SNA, (h-j) noisy limit cycle and (k-m) clean limit cycle.}
    \label{states_35}
\end{figure}

This subset of transitions is observed when the location of $x_f$ is in between $35$ mm and $30$ mm (in figure~\ref{3d_map}\textcolor{blue}{b}, this subset of transitions is plotted with ($\color{blue}\star$) and highlighted in blue). When the location of $x_f = 35$ mm, we observe a continuous transition occurring from a state of intermittency to a state of SNA with a decrease in value of $\phi$ from $2.16$ to $1.59$ (Fig.~\ref{states_35}\textcolor{blue}{a}). A further decrease in the value of $\phi$ results in a shift to a noisy limit cycle with a corresponding abrupt jump of $1100$ Pa in $p'_\mathrm{rms}$. The system further shifts to a state of a clean limit cycle with a continuous increase in $p^\prime_{\mathrm{rms}}$ with a further decrease in the value of $\phi$.\\
\hspace*{10mm}The time series of acoustic pressure oscillations ($p'$) corresponding to the state of intermittency is shown in figure~\ref{states_35}\textcolor{blue}{b}. The corresponding amplitude spectrum has a broad peak of $172$ Hz (Fig.~\ref{states_35}\textcolor{blue}{c}). Decreasing the value of $\phi$ causes a shift to the state of SNA, and $p'$ appear irregular as shown in figure~\ref{states_35}\textcolor{blue}{e}. The amplitude spectrum has two broad peaks at $119$ Hz and $149$ Hz (Fig.~\ref{states_35}\textcolor{blue}{f}). During the state of noisy limit cycle, $p'$ is periodic with amplitude modulations as shown in figure~\ref{states_35}\textcolor{blue}{h}. The corresponding amplitude spectrum has a moderate peak at $143$ Hz (Fig.~\ref{states_35}\textcolor{blue}{i}). Consequently, transitioning to the state of clean limit cycle, the amplitude modulations in $p'$ diminish, leading to a more uniform oscillatory behavior as shown in figure~\ref{states_35}\textcolor{blue}{k} and the dominant frequency becomes $154$ Hz (Fig.~\ref{states_35}\textcolor{blue}{l}).\\
\hspace*{10mm}The trajectory in the reconstructed phase space displays distinct features for each dynamical state. During the state of intermittency, the trajectory alternates between high amplitude periodic and low amplitude aperiodic behavior (Fig.~\ref{states_35}\textcolor{blue}{d}). During the state of SNA, the trajectory appears irregular (Fig.~\ref{states_35}\textcolor{blue}{g}). During the noisy limit cycle, the trajectory forms a thick ring (Fig.~\ref{states_35}\textcolor{blue}{j}), showing minor deviations from the mean limit cycles over a few acoustic cycles. Lastly, during the state of clean limit cycle, the trajectory forms a thin ring structure (Fig.~\ref{states_35}\textcolor{blue}{m}). This subset of transitions may possibly be classified as hybrid or canard types. However, confirming this would require additional experiments with finer resolution in the bifurcation parameter variation, followed by a more detailed analysis. Due to current experimental constraints, these refine variation of the bifurcation parameter is not possible at this time.\\
\begin{table*}
    \caption{\label{table_xf}Summary of the set of transitions obtained for the values of $x_f$ ranges between $70$ mm and $22.5$ mm}
\begin{ruledtabular}
\begin{tabular}{lcccr}
\textbf{Type ($x_f$(mm))} & \textbf{No. of shifts} & \textbf{route of the transition} &  \textbf{ $p'_{rms}$}\\
\hline
Type $1$ ($62.5$)   & $3$ & C$\xrightarrow[\text{}]{\text{cont.}}$ I$\xrightarrow[\text{}]{\text{cont.}}$ SNA $\xrightarrow[\text{}]{\text{cont.}}$ NLC          & \includegraphics[width=1.5 cm,height=0.7 cm]{trans_1.jpg} \\ 
Type $2$ ($45$)     & $4$ & C$\xrightarrow[\text{}]{\text{cont.}}$ I$\xrightarrow[\text{}]{\text{cont.}}$ SNA$\xrightarrow[\text{}]{\text{cont.}}$ NLC $\xrightarrow[\text{}]{\text{cont.}}$ CLC      & \includegraphics[width=1.5 cm,height=0.7 cm]{trans_2.jpg}\\ 
Type $3$ ($37.5$)        & $4$ & C$\xrightarrow[\text{}]{\text{cont.}}$ I$\xrightarrow[\text{}]{\text{cont.}}$ SNA$\xrightarrow[\text{}]{\text{cont.}}$ NLC $\xrightarrow[\text{}]{\text{discont.}}$ CLC   & \includegraphics[width=1.5 cm,height=0.7 cm]{trans_3.jpg}\\ 
Type $5$ ($35$)     & $3$ & I$\xrightarrow[\text{}]{\text{cont.}}$SNA$\xrightarrow[\text{}]{\text{discont.}}$ NLC $\xrightarrow[\text{}]{\text{cont.}}$ CLC       &\includegraphics[width=1.5 cm,height=0.7 cm]{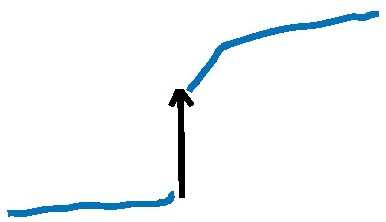}\\
Type $4$ ($22.5$) & $3$ & C$\xrightarrow[\text{}]{\text{cont.}}$ I$\xrightarrow[\text{}]{\text{cont.}}$ SNA $\xrightarrow[\text{}]{\text{discont.}}$ CLC       &\includegraphics[width=1.5 cm,height=0.7 cm]{trans_4.jpg}\\
\end{tabular}
{C-chaos}, {I-intermittency}, {SNA-strange nonchaotic attractor}, {NLC-noisy limit cycle}, {CLC-clean limit cycle}, {cont.-continuous transition} and {discont.-discontinuous transition}
\end{ruledtabular}
\end{table*}

\hspace*{10mm} A summary of the transitions and the corresponding characteristics discussed above is shown in Table~\ref{table_xf}. We discuss the important features observed during the change in the nature of the transition that occurs with varying thermal power input ($\mathcal{P}$) and flame stabilizer position ($x_f$) in Section~\ref{meta}.
 \subsection{The metamorphosis of transition to periodic oscillations} \label{meta}
In figure~\ref{3d_map}, the color code and markers indicate the distinct types of transitions, with arrows representing the abrupt jumps present during the corresponding transitions. The dynamics of the transition undergo a significant transformation with a variation in the secondary parameter. During the transition to periodic oscillations, at the lowest value of $\mathcal{P}$ ($56.5$ kW), the transition occurs in a continuous manner and becomes discontinuous at the highest value of $\mathcal{P}$ ($92.8$ kW). Similarly, at the farthest location of $x_f$ ($70$ mm), the transition occurs in a continuous manner, while at the nearest location of $x_f$ ($22.5$ mm), the nature of the transition changes to discontinuous. Between these two disparate transitions occurring at the extreme values of the secondary parameters, a diverse range of dynamical states exists, changing the dynamics of the transition.\\
\hspace*{10mm}We discuss the commonalities observed within the two sets of transitions. During the observed type $1$ transition, beyond the onset of oscillations, $p'_\mathrm{rms}$ increases gradually with decreasing $\phi$ (the corresponding subset of transitions plotted with $\coloredbox$, highlighted the regime in violet in figure~\ref{3d_map}\textcolor{blue}{a,b}). In contrast, during type $2$ transitions, beyond the onset of oscillations, $p'_\mathrm{rms}$ increases steeply with decreasing $\phi$ (the corresponding subset of transitions highlighted in green in figure~\ref{3d_map}\textcolor{blue}{a,b}). This trend in steep increase in $p'_\mathrm{rms}$ intensifies with increasing $\mathcal{P}$, as illustrated in figure~\ref{3d_map}\textcolor{blue}{a}, and a similar trend in variation of $p'_\mathrm{rms}$ is observed with a variation in $x_f$, as depicted in figure~\ref{3d_map}\textcolor{blue}{b}. This suggests that the nature of the transition gradually evolves from continuous to discontinuous with varying values of the secondary parameter, which can be anticipated through careful examination of the transition. Interestingly, the span of occurrence of type $3$ transition in the secondary parameter space during both sets is minimal.\\
\hspace*{10mm}We discuss the differences observed. During the evolution of the nature of the transition from continuous to discontinuous with increasing $\mathcal{P}$ as a secondary parameter, the onset of the oscillations continues to be delayed till the critical value of $\mathcal{P}$ is reached (respective value of $\phi$ is highlighted with a red box ($\hollowbox{red}$) in each transition (Fig.~\ref{3d_map}\textcolor{blue}{a})). Further increasing in $\mathcal{P}$ beyond the critical value, the onset is advanced (respective value of $\phi$ is highlighted with a blue box ($\hollowbox{blue}$) in each transition (Fig.~\ref{3d_map}\textcolor{blue}{a})).\\
\hspace*{10mm}In contrast, during the evolution of the nature of the transition from continuous to discontinuous with a decreasing value of $x_f$ as a secondary parameter, the onset of oscillations is not significantly affected until the critical value of $x_f$ is reached (respective value of $\phi$ is highlighted with a red box ($\hollowbox{red}$) in each transition (Fig.~\ref{3d_map}\textcolor{blue}{b})). With a further decrease in the value of $x_f$, the onset advances (respective value of $\phi$ is highlighted with a blue box ($\hollowbox{blue}$) in each transition (Fig.~\ref{3d_map}\textcolor{blue}{b})). 
\section{Discussion and conclusion}\label{discussion}
In our study, we perform experiments in a turbulent reactive flow system with a bluff body to stabilize the flame and obtain a transition from chaotic fluctuations to periodic oscillations, decreasing the equivalence ratio as the bifurcation parameter. We also vary a secondary parameter to examine the effect on the nature of this transition. During this transition, the system undergoes dynamical shifts between various dynamical states, such as the states of chaos,  intermittency, strange nonchaotic attractor (SNA), noisy limit cycle, and clean limit cycle. \\
\hspace*{10mm}We find that the transition to periodic oscillations can be either continuous or discontinuous, depending on the secondary parameters, such as the thermal power input or the location of the flame stabilizer position. We discover that the nature of the transition changes from continuous to discontinuous by systematically increasing the value of thermal power input or decreasing the location of the flame stabilizer position individually in combination with a quasi-static increase in the bifurcation parameter.\\
\hspace*{10mm}A change in the nature of the transition from super critical Hopf bifurcation to subcritical Hopf bifurcation was reported in a laminar thermoacoustic system by \citet{etikyala2017change}. They demonstrated that at lower airflow rates, increasing the bifurcation parameter leads to a supercritical Hopf bifurcation, whereas at higher airflow rates, a subcritical Hopf bifurcation was observed. A recent mathematical investigation by \citet{kuehn2021universal} showed that a universality exists in the change in the nature of the transition from continuous to discontinuous. They made this argument by mathematically analyzing three dynamical systems where they varied a secondary parameter, thereby inducing higher-order nonlinear effects during the transition.\\
\hspace*{10mm}Interestingly, in the transition to periodic oscillations within turbulent combustors, the system traverses through more intricate dynamical states beyond just fixed points and limit cycles. This complexity suggests that a simple Hopf bifurcation model may not fully capture the intricate dynamics involved in the transition to thermoacoustic instability in turbulent reactive flows. Turbulence encompasses a broad spectrum of eddies spanning from very small to very large scales. As a result, turbulence enhances local interactions across different spatial locations within the combustion field.\\
\hspace*{10mm}The increase in the value of thermal power input increases not only does the energy in the system rise, but there is also a corresponding need for increased airflow to maintain the equivalence ratio, which intensifies turbulence. This enhanced turbulence increases local-scale interactions, while the rise in energy input strengthens interactions among the subsystems. These phenomena might be impacting the transition and causing a change in the nature of the transition from continuous to discontinuous. Varying the location of the flame stabilizer position affects the local velocity, thereby influencing the frequency of the vortex shedding from the tip of the flame stabilizer.\\
\hspace*{10mm}Therefore, the change in the nature of the transition is more intricate in turbulent reactive flow systems and undergoes a metamorphosis of the transition. Spatiotemporal analysis of data, such as high-speed chemiluminescence images and local velocity and vorticity fields obtained using particle image velocimetry, could provide valuable insights into (a) the characteristic of each dynamical state observed, (b) the distinct transitions identified, and thereby (c) the metamorphosis of a transition in turbulent reactive flows.
\begin{acknowledgments}
We thank S. Anand, S. Thilagaraj, and G. Sudha for their help during the experiments. T. B. would like to extend sincere gratitude to Dr. Induja Pavithran for their invaluable guidance and mentorship. T. B. acknowledges the research assistantship from the Ministry of Human Resource Development, India, and the Indian Institute of Technology Madras. R. I. S. acknowledges the funding from the Science and Engineering Research Board (SERB) of the Department of Science and Technology (DST) through a J. C. Bose Fellowship (No. JCB$/2018/000034/$SSC) and from the IOE initiative (No. SP$22231222$CPETWOCTSHOC).
\end{acknowledgments}
\section*{Data Availability Statement}
The data that support the findings of this study are available from the corresponding author upon reasonable request.
\appendix 
\section*{appendix}
In Fig.~\ref{bif_bb} we present the bifurcation diagrams corresponding to the four types of transitions observed with varying the location of flame stabilizer position ($x_f$) as a secondary parameter. Following this, we provide plots of a detailed analysis of the time series for the dynamical states observed during each transition in Appendix~\ref{xf_BB}. 
\section{Time series analysis of four types of transitions with flame stabilizer position} \label{xf_BB}
\begin{figure}
     \centering
    \includegraphics[width= 8.5 cm,height= 6.5 cm]{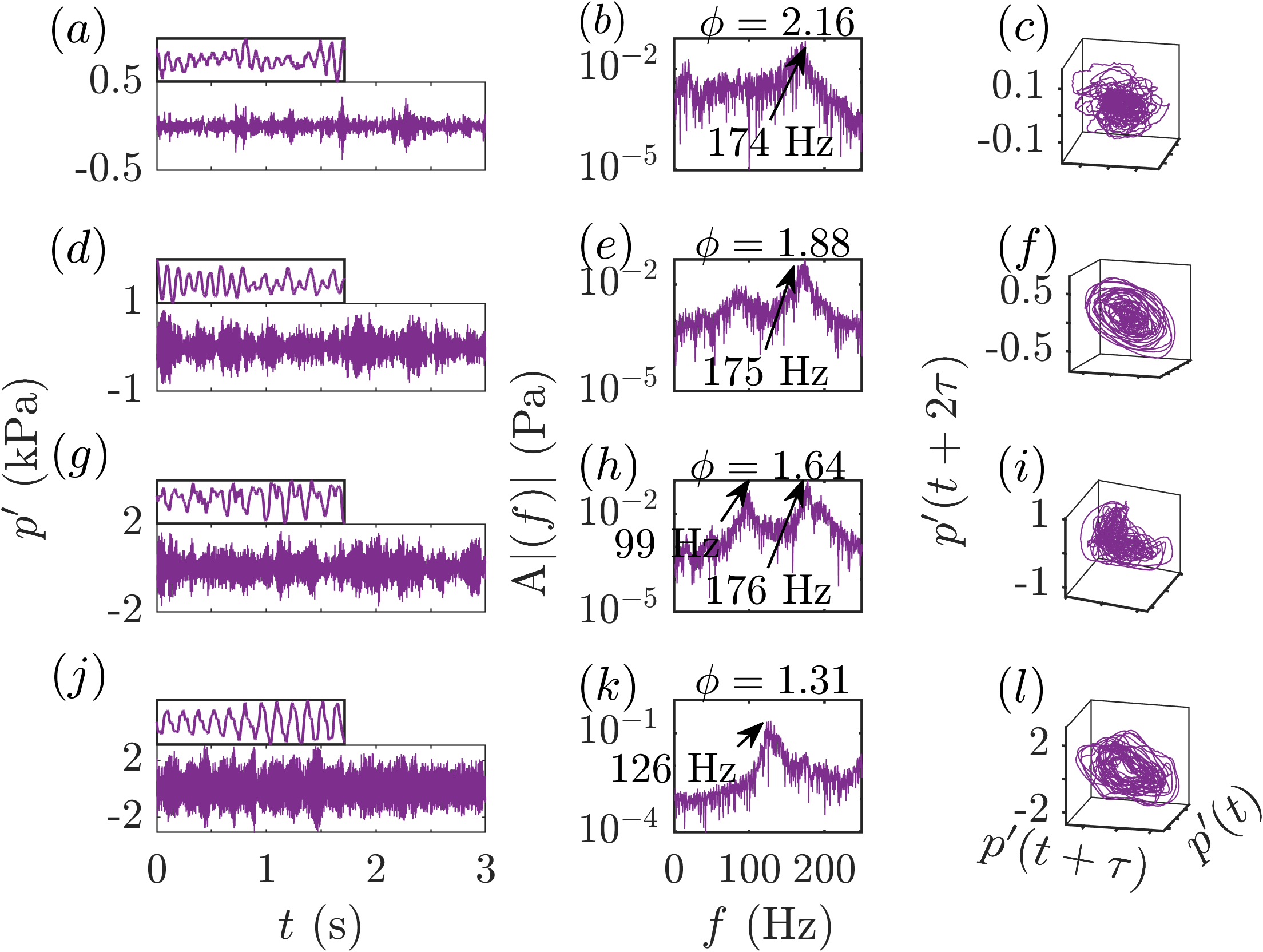}
    \caption{\textbf{Type $1$: C-I-SNA-NLC route of continuous transition}. Time series, amplitude spectrum, and the reconstructed phase space of $p'$ during the states of (a-c) chaos, (d-f) intermittency, (g-i) SNA, and (j-l) noisy limit cycle, observed during the bifurcation shown in figure~\ref{bif_bb}\textcolor{blue}{a}.}
    \label{states_62.5}
 \end{figure}

The characteristics of the time series of acoustic pressure fluctuations ($p'$) during the observed dynamical states are qualitatively consistent with those presented in Sections~\ref{56.5kw} and \ref{35mm}. During type $1$ transition systems transitions from chaotic fluctuations to noisy limit cycle oscillations traversing via states of intermittency and SNA (Fig.~\ref{states_62.5}). We observe a significant frequency shift during the bifurcation from state of SNA to noisy limit cycle.\\

 \begin{figure}
     \centering
    \includegraphics[width= 8.5 cm,height= 10 cm]{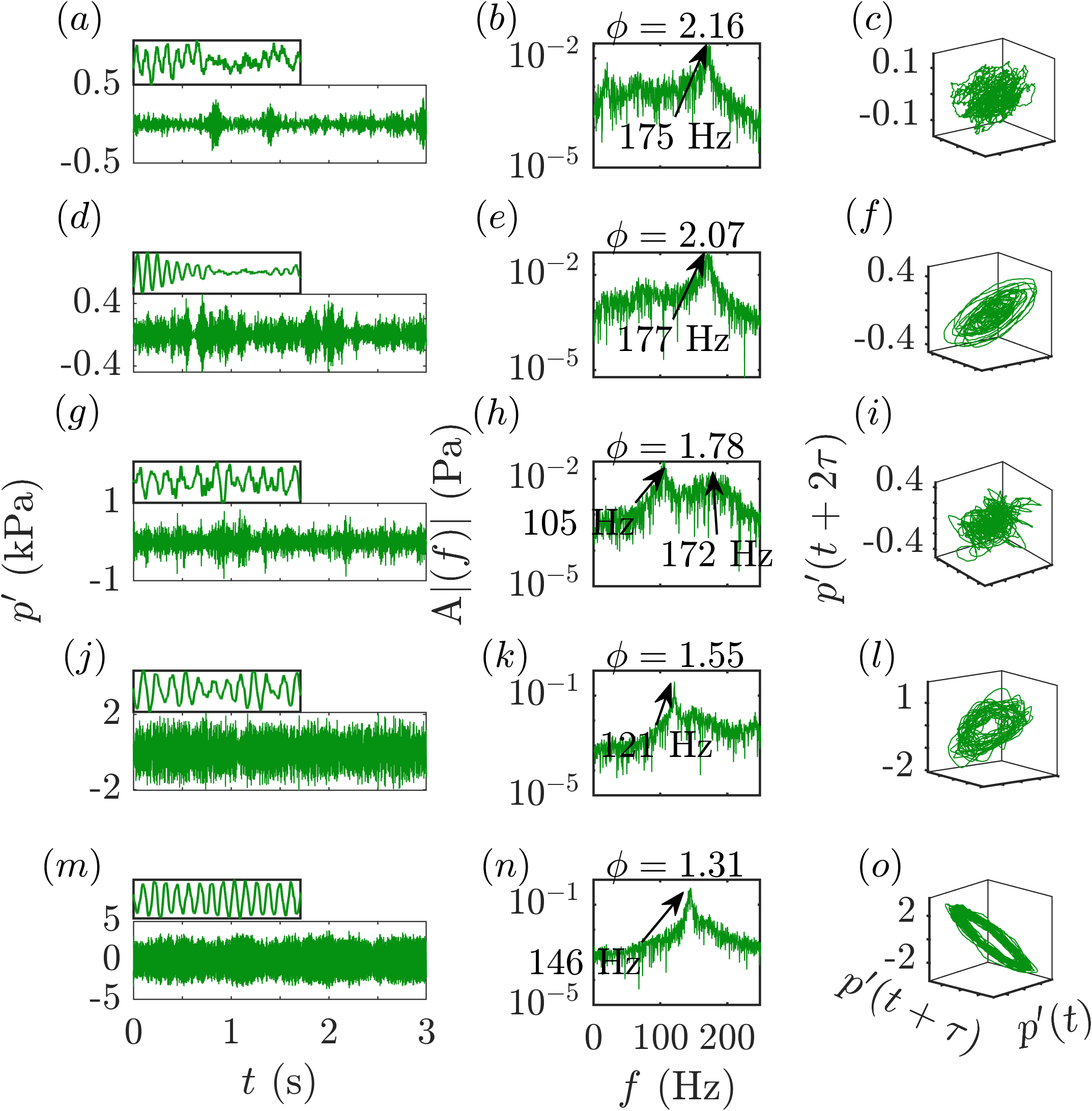}
    \caption{\textbf{Type $2$: C-I-SNA-NLC-CLC route of continuous transition}. Time series, amplitude spectrum, and the reconstructed phase space of $p'$ during the states of (a-c) chaos, (d-f) intermittency, (g-i) SNA, (j-l) noisy limit cycle, (m-o) clean limit cycle, observed during the bifurcation shown in figure~\ref{bif_bb}\textcolor{blue}{b}.}
    \label{states_45}
 \end{figure}
 \hspace*{10mm}In type $2$ transition, the system transitions from chaotic fluctuations to clean limit cycle oscillations and traversed via states of intermittency, SNA, and noisy limit cycle (Fig.~\ref{states_45}). In this transition, the system obtained high amplitude oscillations compared to type $1$ transition.\\ 

 \begin{figure}
     \centering
    \includegraphics[width= 8.5 cm,height= 10 cm]{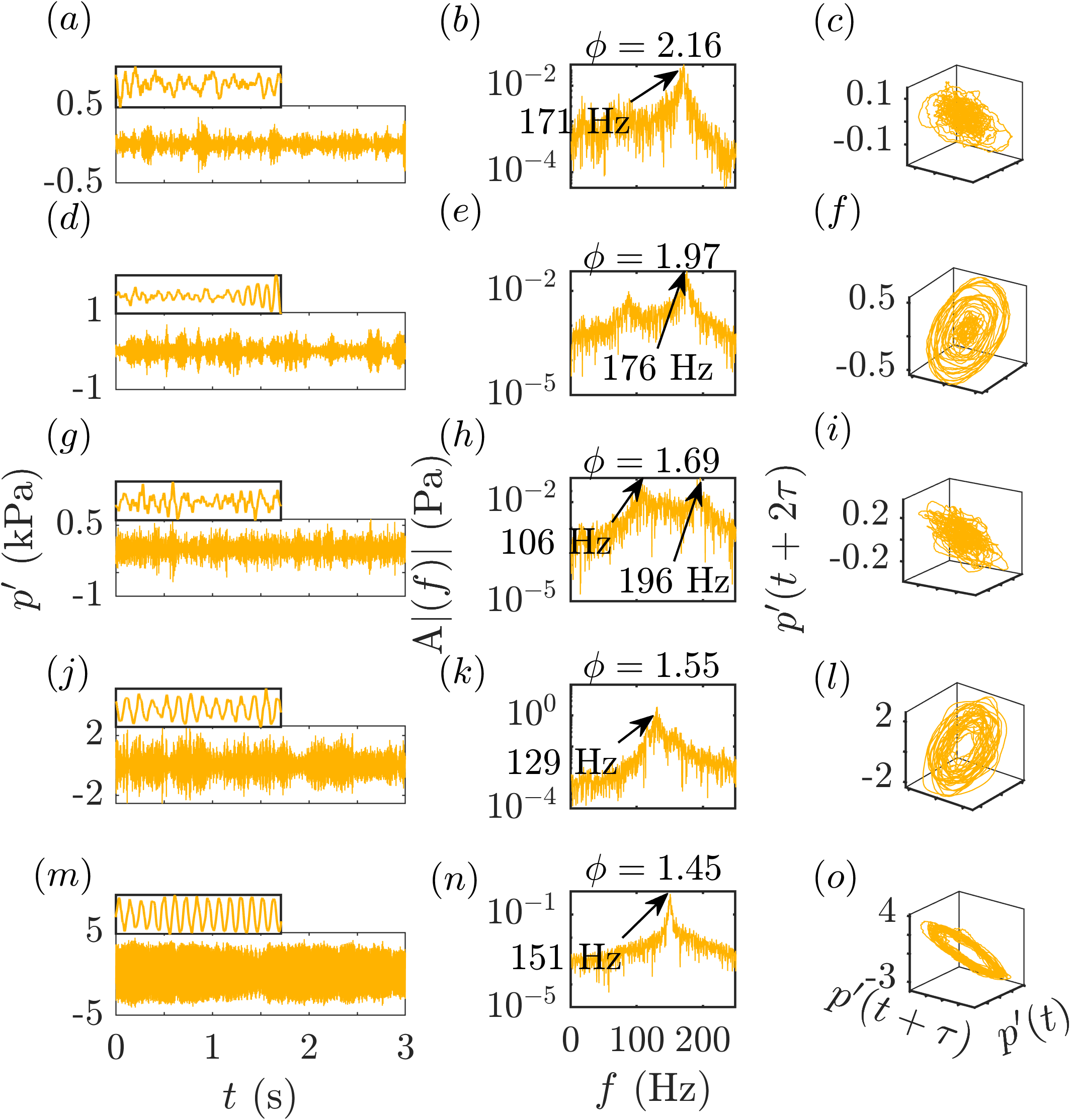}
    \caption{\textbf{Type $3$: C-I-SNA-NLC-CLC route of primary continuous bifurcation to NLC followed by a secondary discontinuous bifurcation to CLC}. Time series, amplitude spectrum, and the reconstructed phase space of $p'$ during the states of (a-c) chaos, (d-f) intermittency, (g-i) SNA, (j-l) noisy limit cycle, (m-o) clean limit cycle, observed during the bifurcation shown in figure~\ref{bif_bb}\textcolor{blue}{c}.}
    \label{states_37.5}
 \end{figure}
\hspace*{10mm}During type $3$ transition the system undergo a continuous bifurcation to noisy limit cycle, followed by an abrupt transition to a clean limit cycle (Fig.~\ref{states_37.5}). This abrupt jump associates with a frequency shift of $22$ Hz.\\
 
 \begin{figure}
    \centering
    \includegraphics[width = 8.5 cm, height = 8cm]{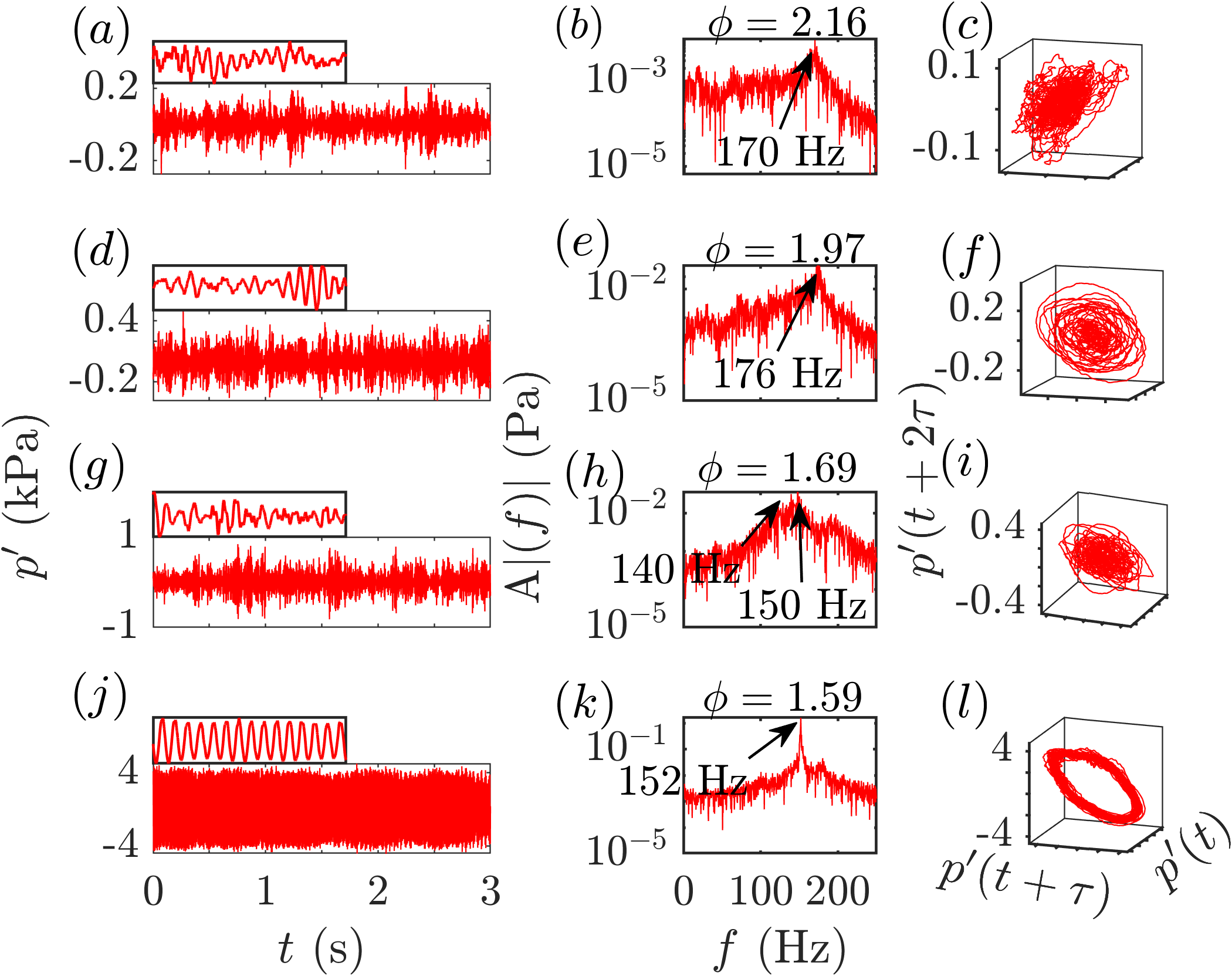}
    \caption{\textbf{Type $4$: C-I-SNA-CLC route of discontinuous transition}. Time series, amplitude spectrum, and the reconstructed phase space of $p'$ during the states of (a-c) chaos, (d-f) intermittency, (g-i) SNA, and (j-l) clean limit cycle oscillations, observed during the bifurcation shown in figure~\ref{bif_bb}\textcolor{blue}{d}.}
    \label{states_22.5}
\end{figure}
\hspace*{10mm}Finally, during type $4$ transition, the system abruptly transition to clean limit cycle oscillations from state of SNA (Fig.~\ref{states_22.5}). Here, the system skipped the state of noisy limit cycle.
 
\nocite{*}
\clearpage
\bibliography{apssamp}

\end{document}